\documentclass{amsart}
\usepackage{graphicx,amsmath,cite}
\newtheorem{thm}{Theorem}[section]

\newtheorem{lem}[thm]{Lemma}

\theoremstyle{definition}

\theoremstyle{remark}

\numberwithin{equation}{section}

\newcommand{\mb}[1]{\mathbb{#1}}
\def\beq{\begin{eqnarray}}
\def\eeq{\end{eqnarray}}
\begin{document}
\title[Bianchi Cosmologies: A Tale of Two Tilted fluids]{Bianchi Cosmologies: \\ A Tale of Two Tilted fluids}

\author{Alan A. Coley \& Sigbj{\o}rn Hervik}%
\address{Department of Mathematics \& Statistics, Dalhousie University,
Halifax, Nova Scotia,
Canada B3H 3J5}%
\email{aac@mathstat.dal.ca, herviks@mathstat.dal.ca}%

\subjclass{}%
\keywords{}%

\date{\today}%
\begin{abstract}
We use a dynamical systems approach to study Bianchi type VI$_0$
cosmological models containing two tilted $\gamma$-law perfect
fluids. The full state space is 11-dimensional, but the existence
of a monotonic function simplifies the analysis considerably. We
restrict attention to a particular, physically interesting,
invariant subspace and find all equilibrium points that are future
stable in the full 11-dimensional state space; these are
consequently local attractors and serve as late-time asymptotes
for an open set of tilted type VI$_0$ models containing two tilted
fluids. We find that if one of the fluids has an equation of state
parameter $\hat{\gamma}<6/5$, the stiffest fluid will be
dynamically insignificant at late times. For the value
$\hat\gamma=6/5$ there is a 2-dimensional bifurcation set, and if
both fluids are stiffer than $\hat\gamma=6/5$  both fluids will
have extreme tilt asymptotically. We investigate the case in which
one fluid is extremely tilting in detail. We also consider the
case with one stiff fluid ($\hat\gamma=2$) close to the initial
singularity, and find that the  chaotic behaviour which occurs in
general Bianchi models with $\hat\gamma<2$ is suppressed.
\end{abstract}
\maketitle

\section{Introduction}

In \cite{hervik} the asymptotic late-time behaviour of general
tilted Bianchi type VI$_0$ universes with a tilted $\gamma$-law
perfect fluid was analysed. All of the future stable equilibrium
points for various subclasses of tilted type VI$_0$ models, as well
as for the general tilted type VI$_0$ models, were found. In particular,
for the particular value of the equation of state parameter,
$\gamma=6/5$, there exists a bifurcation line which signals a
transition of stability between a non-tilted equilibrium point to
an extremely tilted equilibrium point, confirming the observation
in \cite{BHtilted}, which followed up on the earlier work on
tilted fluids \cite{KingEllis,BN,BS,HBWII,Shikin,Collins,HWV}.

In this paper, we will generalize this work and use the dynamical
systems approach \cite{WE,COLEY2} to analyse the Bianchi type
VI$_0$ model containing two (in general both) tilted (and
separately conserved) $\gamma$-law perfect fluids \cite{CW}. We
shall use the formalism employed in \cite{hervik}, and pay
particular attention to some of the cases which were not analyzed
in detail there. In particular, we shall study the two physically
important cases of when the second fluid is a comoving stiff fluid
(i.e., $\Gamma=2$ and $\tilde{V}=0$ in the notation below), which represents a
scalar field close to the singularity, and when the second fluid
is of extreme tilt (i.e., $\tilde{V}=1$). The study of extremely tilted fluids is partly motivated from
the brane-world scenario where gravitational waves in the bulk can, through the interaction of the bulk
Weyl tensor with the brane, give rise to extremely tilted fluids as seen from a brane point of view.

The Bianchi type VI$_0$ model  is not the most general model but is
sufficiently general to account for many interesting phenomena.
Bianchi type VI$_0$ models with various sources of matter has been
studied. The type VI$_0$ model with a comoving perfect fluid was
considered in \cite{WH:89}, with a magnetic and a $\gamma$-law
perfect fluid in \cite{LKW}, and with a tilted $\gamma$-law
perfect fluid in \cite{hervik}. In all of these studies there has
been evidence that the type VI$_0$ model is asymptotically self-similar
at late times; the generic solution will approach an equilibrium
point in the dynamical system. In the two tilted model the state
space is 11-dimensional, and obtaining all of the equilibrium
points is an extremely laborious task. However, guided by 
previous work we concentrate on the equilibrium points in a
particularly interesting subspace; we will then show that
these equilibrium points are stable in the \emph{full
11-dimensional state space}.

The paper is organised as follows: Next, in section 2, we write
down the equations of motion using the orthonormal frame formalism
and discuss the various invariant subspaces. In section
\ref{sect:latetime} we discuss the (relevant) equilibrium points
and their stability, and present some general results regarding
the asymptotic behaviour of the tilted type VI$_0$ model. In
section \ref{sect:extreme} we consider the case in which one fluid
is extremely tilted. In section \ref{sect:stiffone} we consider
the case of a stiff fluid. Close to the initial singularity a
scalar field is essentially massless and can be modelled by a
comoving perfect fluid with a stiff equation of state parameter
$\gamma=2$. This will enable us to consider the initial singular
regime in this case. Finally we  summarise our results  in section
\ref{sect:disc}.

\subsection{Preliminaries}

In this paper we shall assume the existence of two (in general
both tilting) perfect fluids. We shall denote the second fluid
using tildes (e.g., $\tilde{\Omega}, \tilde{V}$), and use the
notation $\Gamma=\tilde{\gamma}$, where by convention we assume
that $\Gamma>\gamma$. We shall also assume that each fluid is
separately conserved.

The energy-momentum tensor for each tilted perfect fluid is \beq
T_{\mu\nu}=(\hat\rho +\hat{p})\hat{u}_{\mu}\hat{u}_{\nu}+\hat
pg_{\mu\nu}, \eeq where $\hat{u}^{\mu}=(\cosh\theta,\sinh\theta
c^a)$ is the fluid velocity and $\theta$ its rapidity. The spatial vector $c^a$ is chosen to
be a unit vector in the tangent space of the surfaces of homogeneity; i.e. $c^ac_a=1$. We will further
assume that the each fluid obeys the barotropic equation of state,
\beq \hat p=(\hat{\gamma}-1)\hat\rho. 
\eeq 
In terms of the unit
normal vector ${\bf u}={\bf e}_0$ to the group orbits the
energy-momentum tensor takes the imperfect fluid form, where
$\rho, p, q_{a}, \pi_{ab}$ can be explicitly written down. For the
Bianchi cosmologies  we can always write the line-element in
canonical form with respect to the structure constants
$C_{~bc}^{a}$ of the simply transitive Bianchi group type under
consideration, which depend only on time \cite{EM}. For the type
VI$_0$ model, $a_c=0$ and $n_{ab}$ has two non-zero eigenvalues
with opposite sign. This implies that we can choose a frame such
that the structure constants can be written \beq
n_{ab}=\begin{bmatrix}
0 &0 & 0 \\
0 & \bar{n} & n \\
0 & n & \bar{n}
\end{bmatrix}, \qquad a_b=0.
\eeq Furthermore, the type VI$_0$ has $\bar{n}^2<n^2$. The
equations of motion can now be written down in terms of
expansion-normalised variables.


\section{Equation of motion}
We define the expansion-normalised variables:
\beq \Sigma_{ab}=\begin{bmatrix}
-2\Sigma_+ & \sqrt{3}\Sigma_{12} & \sqrt{3}\Sigma_{13} \\
\sqrt{3}\Sigma_{12} & \Sigma_++\sqrt{3}\Sigma_- &
\sqrt{3}\Sigma_{23}
\\
 \sqrt{3}\Sigma_{13} & \sqrt{3}\Sigma_{23} & \Sigma_+-\sqrt{3}\Sigma_-
\end{bmatrix},  \\
N_{22}=N_{33}=\sqrt{3}\bar{N},\quad N_{23}=N_{32}=\sqrt{3}N.
\eeq 
We also introduce the three-velocity $V$ by \beq \sinh\theta
=\frac{V}{\sqrt{1-V^2}},\quad 0\leq V<1. \eeq
It is also
convenient  to introduce a parameter $\lambda$ defined by
$\bar{N}=\lambda N$, which makes it possible to solve for the
function $R_1$ (where $R_a$ is the local angular velocity of a Fermi-propagated axis with respect to the triad ${\bf e}_a$) and eliminate it completely from the equations
of motion. For the Bianchi type VI$_0$ model this parameter will
be bounded; i.e. $\lambda^2<1$.

For the expansion-normalised variables the equations of motion
are: \beq \Sigma_+'&=&
(q-2)\Sigma_++{3}(\Sigma_{12}^2+\Sigma^2_{13})-2N^2 \nonumber \\
&&
+\frac{\gamma\Omega}{2G_+}\left(-2v_1^2+v_2^2+v_3^2\right)+\frac{\Gamma\tilde{\Omega}}{2\tilde{G}_+}\left(-2\tilde{v}_1^2+\tilde{v}_2^2+\tilde{v}_3^2\right) \\
\Sigma_-'&=&
(q-2)\Sigma_-+\sqrt{3}(\Sigma_{12}^2-\Sigma_{13}^2)-2\sqrt{3}\lambda\Sigma_-\Sigma_{23}\nonumber
\\ &&
+\frac{\sqrt{3}\gamma\Omega}{2G_+}\left(v_2^2-v_3^2\right)-\frac{\sqrt{3}\Gamma\tilde{\Omega}}{2\tilde{G}_+}\left(\tilde{v}_2^2-\tilde{v}_3^2\right)
\\
\Sigma'_{12}&=&
\left(q-2-3\Sigma_+-\sqrt{3}\Sigma_-\right)\Sigma_{12}
-\sqrt{3}\left(\lambda\Sigma_-+\Sigma_{23}\right)\Sigma_{13} \nonumber \\
&&+\frac{\sqrt{3}\gamma\Omega}{G_+}v_1v_2+\frac{\sqrt{3}\Gamma\tilde\Omega}{\tilde{G}_+}\tilde{v}_1\tilde{v}_2
\\
\Sigma'_{13}&=&\left(q-2-3\Sigma_++\sqrt{3}\Sigma_-\right)\Sigma_{13}
 -\sqrt{3}\left(-\lambda\Sigma_-+\Sigma_{23}\right)\Sigma_{12} \nonumber \\
 &&+\frac{\sqrt{3}\gamma\Omega}{G_+}v_1v_3+\frac{\sqrt{3}\Gamma\tilde\Omega}{\tilde{G}_+}\tilde{v}_1\tilde{v}_3
\\
\Sigma'_{23}&=&(q-2)\Sigma_{23}-2\sqrt{3}\lambda
N^2+2\sqrt{3}\lambda\Sigma_-^2+2\sqrt{3}\Sigma_{12}\Sigma_{13}\nonumber
\\
&& +
\frac{\sqrt{3}\gamma\Omega}{G_+}v_2v_3+\frac{\sqrt{3}\Gamma\tilde\Omega}{\tilde{G}_+}\tilde{v}_2\tilde{v}_3
\\
N'&=& \left(q+2\Sigma_++2\sqrt{3}\Sigma_{23}\lambda\right){N}\\
\lambda'&=&2\sqrt{3}\Sigma_{23}(1-\lambda^2)\label{eq:lambda}
\eeq The equations for the fluids are (similar equations for the
second fluid):
\beq \Omega'&=&
\frac{\Omega}{G_+}\Big\{2q-(3\gamma-2)
 +\left[2q(\gamma-1)-(2-\gamma)-\gamma\mathcal{S}\right]V^2\Big\}
 \quad \\
 v_1' &=& \left(T+2\Sigma_+\right)v_1-2\sqrt{3}\Sigma_{13}v_3-2\sqrt{3}\Sigma_{12}v_2\nonumber \\
 && -\sqrt{3}N\left(v_2^2-v_3^2\right)\\
 v_2'&=& \left(T-\Sigma_+-\sqrt{3}\Sigma_-\right)v_2-\sqrt{3}\left(\Sigma_{23}+\lambda\Sigma_-\right)v_3\nonumber \\
 && +\sqrt{3}\lambda{N}v_1v_3+\sqrt{3}Nv_1v_2 \\
 v_3'&=& \left(T-\Sigma_++\sqrt{3}\Sigma_-\right)v_3-\sqrt{3}\left(\Sigma_{23}-\lambda\Sigma_-\right)v_2\nonumber \\
 && -\sqrt{3}\lambda{N}v_1v_2-\sqrt{3}Nv_1v_3 \\
  V'&=&
\frac{V(1-V^2)}{1-(\gamma-1)V^2}\left[(3\gamma-4)-\mathcal{S}\right],
\eeq where \beq q&=& 2\Sigma^2+\frac
12\frac{(3\gamma-2)+(2-\gamma)V^2}{1+(\gamma-1)V^2}\Omega+\frac
12\frac{(3\Gamma-2)+(2-\Gamma)\tilde{V}^2}{1+(\Gamma-1)\tilde{V}^2}\tilde{\Omega}\nonumber \\
\Sigma^2 &=& \Sigma_+^2+\Sigma_-^2+\Sigma_{12}^2+ \Sigma_{13}^2+\Sigma_{23}^2\nonumber \\
\mathcal{S} &=& \Sigma_{ab}c^ac^b, \quad c^ac_{a}=1, \quad v^a=Vc^a,\quad \nonumber \\
 V^2 &=& v_1^2+v_2^2+v_3^2,\quad  \nonumber \\
G_+ &=& 1+(\gamma-1)V^2, \nonumber \\
 T&=& \frac{(3\gamma-4)(1-V^2)+(2-\gamma)V^2\mathcal{S}}{1-(\gamma-1)V^2}.
\eeq 
These variables are subject to the constraints 
\beq
1&=& \Sigma^2+N^2+\Omega +\tilde{\Omega}\label{const:H}\\
0 &=& 2\Sigma_-N+\frac{\gamma\Omega v_1}{G_+}+\frac{\Gamma\tilde{\Omega}\tilde{v}_1}{\tilde{G}_+} \label{const:v1}\\
0 &=&
-\left(\Sigma_{12}+\Sigma_{13}\lambda\right)N+\frac{\gamma\Omega v_2}{G_+}+\frac{\Gamma\tilde{\Omega}\tilde{v}_2}{\tilde{G}_+} \label{const:v2}\\
0 &=&
\left(\Sigma_{13}+\Sigma_{12}\lambda\right)N+\frac{\gamma\Omega
v_3}{G_+}+\frac{\Gamma\tilde{\Omega}\tilde{v}_3}{\tilde{G}_+}.
\label{const:v3}
\eeq 
The parameter $\gamma$ will be assumed to be in the interval
$\gamma\in ( 0,2]$ and we can without loss of
generality assume that $\gamma<\Gamma$.

Eqs (\ref{const:v1})-(\ref{const:v3}) essentially define $\tilde{v}_i$, and so the
governing equations are a twelve-dimensional set subject to a
single constraint (eq. (\ref{const:H})); hence the dynamical system is
effectively eleven-dimensional.

\subsection{The state space} The
constraint (\ref{const:H}) implies that
$\Sigma_{\pm},~\Sigma_{12},~\Sigma_{13},~\Sigma_{23},~N$, $\Omega$
and $\tilde{\Omega}$
 are all bounded. Combining the equations for $N$ and $\bar{N}$ we get the
equations
\beq \left(N\pm \bar{N}\right)'&=& \left(q+2\Sigma_+\pm
2\sqrt{3}\Sigma_{23}\right)\left(N\pm \bar{N}\right).
\eeq
Thus
if the initial data have $\bar{N}^2<N^2$, then this will hold for
all times\footnote{This can also be seen by using $\lambda$ and eq.(\ref{eq:lambda}): $\bar{N}^2<N^2\Leftrightarrow \lambda^2<1$.}.
Hence, for type VI$_0$, $\bar{N}$ will be bounded as
well. The invariant subspaces $N\pm\bar{N}=0$ correspond to
Bianchi type II universes. We also require $0\leq V,
\tilde{V}\leq 1$ for physical reasons. This implies the bounds
\beq \Sigma_+^2+\Sigma_-^2+\Sigma_{12}^2+
\Sigma_{13}^2+\Sigma_{23}^2+N^2
&\leq & 1 \nonumber \\
\bar{N}^2 & \leq  & N^2. \nonumber  \label{bounds}\eeq 
We have,
therefore, the 15 bounded variables
$\Sigma_{\pm},~\Sigma_{12},~\Sigma_{13},~\Sigma_{23},~N$,
$\bar{N}$, $v_i$, $\tilde{v}_i$, $\Omega$ and $\tilde{\Omega}$.
However, due to 4 constraints, these are not all independent.
Hence, the state space can be considered a subspace of a compact
region in $\mb{R}^{11}$. There are also some discrete symmetries
which are intrinsic to the type VI$_0$ geometry \cite{hervik}.

\subsection{Invariant subspaces} \label{sect:inv}

Some of the physically interesting invariant subspaces are as
follows (we will also assume that the boundaries are included):
\begin{enumerate}
\item{} $T(VI_0)$: The full state space of tilted type VI$_0$.
\item{} $F(VI_0)$: $\Sigma_-=0$, $\Sigma_{12}=-\Sigma_{13}$, $v_1=0$, $v_2=v_3$.
\item{} $T_2(VI_0)$: A Bianchi tilted type VI$_0$ with two tilt-degrees of freedom; $\Sigma_{12}=\Sigma_{23}=\bar{N}=0$ (or
$\Sigma_{13}=\Sigma_{23}=\bar{N}=0$), $v_2=0$.
\item{} $T_1(VI_0)$: Tilted Bianchi
type VI$_0$ with one tilt degree of freedom;
$\Sigma_{12}=\Sigma_{13}=0$, $v_2=v_3=0$.
\item{} $B(VI_0)$: Non-tilted Bianchi
type VI$_0$; $\Sigma_-=\Sigma_{12}=\Sigma_{13}=V=\tilde{V}=0$.
\item{}
$T^{\pm}(II)$: The tilted type II boundary; $N=\pm \bar{N}$.
\item{} $B(II)$: Non-tilted type II; $N^2=\bar{N}^2$ and $V=\tilde{V}=0$.
\item{} $B(I)$: Bianchi type I universes; $N=\bar{N}=V=\tilde{V}=0$.
\item{}
$\partial_{V} B(I)$: The Bianchi type I vacuum boundary;
$N=\bar{N}=V=\Omega=\tilde{V}=\tilde{\Omega}=0$.
\end{enumerate}

\subsection{A monotonic function} Fortunately, the above system of
equations possess a monotonic function with makes it possible to
determine an important result regarding the asymptotic behaviour
of the two-fluid model. We define the two functions: \beq
\beta=\frac{(1-V^2)^{\frac 12(2-\gamma)}}{1+(\gamma-1)V^2}, \quad
\tilde{\beta}=\frac{(1-\tilde{V}^2)^{\frac
12(2-\Gamma)}}{1+(\Gamma-1)\tilde{V}^2}. \label{eq:betas}\eeq Then
the function, \beq \chi\equiv
\frac{\beta\Omega-\tilde{\beta}\tilde{\Omega}}
{\beta\Omega+\tilde{\beta}\tilde{\Omega}}, \eeq where $\chi$ is
bounded and satisfies $-1 \le \chi \le 1$, satisfies the
differential equation \beq \chi'=\frac
32(\Gamma-\gamma)(1-\chi^2). \eeq Hence, assuming $\Gamma>\gamma$,
then $\chi' \ge 0$ and $\chi$ is monotonically increasing. From
this we can deduce that (for $\chi^2\neq 1$)
\begin{enumerate}
\item{} $\lim_{\tau\rightarrow-\infty}\chi=-1$; i.e., at early
times $\beta\Omega \rightarrow 0$, \item{}
$\lim_{\tau\rightarrow+\infty}\chi=+1$; i.e., at late times
$\tilde{\beta}\tilde{\Omega}\rightarrow 0$.
\end{enumerate}

At early times, we have that $\beta\Omega \rightarrow 0$. If
$\beta \ne 0$ (i.e., not extreme tilt), it follows that $\Omega
\rightarrow 0$ and the first (less stiff) fluid is dynamically
negligible at early times. Indeed, if neither fluids have extreme
tilt asymptotically, then $\lim_{\tau\rightarrow-\infty}\Omega=0$,
and $\lim_{\tau\rightarrow+\infty}\tilde{\Omega}=0$. Hence,
asymptotically the dynamical behaviour is that of a single fluid
cosmology. It is therefore of interest to study the behaviour for
when (at least) one of the fluids have extreme tilt and under what
circumstances this can occur asymptotically. Note that in the case
of a comoving stiff perfect fluid ($\tilde{\beta}=1$ and $\Gamma
=2$), the future asymptotic behaviour is governed by the first
fluid (and is given in \cite{hervik}), and the behaviour close to
the initial singularity need not be oscillatory.

The behaviour of the monotonic function $\chi$ combined with the
earlier work \cite{hervik} also gives us a hint of where to find
the relevant equilibrium points. In \cite{hervik} we noticed
that all the locally future stable equilibrium points occurred in
the invariant subspace $F(VI_0)$. The monotonic function $\chi$
tells us that the late-time behaviour is almost entirely dictated
by the less stiff fluid. Hence, one would expect that all the
locally future stable  equilibrium points for the two-fluid model
also lie in this invariant subspace. Since the total state space
is 11-dimensional, this observation simplifies our analysis
considerably. In the following we shall therefore only consider
equilibrium points in $F(VI_0)$, but the stability analysis will
be performed in the full 11-dimensional state space.

\section{Late-time behaviour}
\label{sect:latetime} The system of equations possesses a wealth
of equilibrium points. In Appendix \ref{sect:Eqpoints} we have
listed all of the physically interesting ones in the invariant
subspace $F(VI_0)$.

By inspection of the equilibrium points and their eigenvalues, we
see that they all play a role in the late-time behaviour for
various values of the equation of state parameters. Based on the
analysis in the Appendix, the equilibrium points that are locally
stable to the future are summarized in Table \ref{tab:stable};
these represent the possible future asymptotic states of the
models. It should be emphasised that some of these equilibrium points have zero eigenvalues so when we refer to stability we mean in the technical sense; e.g., there is 1-dimensional equilibrium set which is normally hyperbolic.

There is another thing worth mentioning regarding the second
fluid. If $\tilde{\Omega}\rightarrow 0$ , then the second fluid
will decouple and become dynamically insignificant. For these
cases the tilt velocity $\tilde{V}$ of the equilibrium points
indicates the tilt velocity close to the equilibrium points. Only
away from (but arbitrary close to) $\tilde{\Omega}=0$ is
$\tilde{V}$ a meaningful quantity.

Note that for $\gamma=6/5$ there is a 2-dimensional bifurcation
set and for $\gamma\geq 6/5$ both of the fluids are dynamically
significant at late times. If $\gamma <6/5$, on the other hand,
the stiffest fluid becomes dynamically insignificant and the late
time asymptotic behaviour of the model is entirely dominated by
the less stiff fluid.

\begin{table}
\centering
\begin{tabular}{|c|c|c|}
\hline $\gamma$ & $\Gamma$ & Future Attractor\\ \hline \hline  &
$\gamma <\Gamma <4/3$ & $\mathcal{I}(I)$
\\  $0<\gamma < 2/3$ & $\Gamma =4/3$ & $\mathcal{I}_T(I)$ \\
 & $4/3 <\Gamma $ & $\mathcal{I}_E(I)$ \\ \hline
   & $\gamma <\Gamma <4/3-(3\gamma-2)/12$ & $\mathcal{C}(VI_0)$ \\  $2/3<\gamma
< 6/5$ & $\Gamma =4/3-(3\gamma-2)/12$ & $\mathcal{C}_T(VI_0)$ \\
 & $4/3-(3\gamma-2)/12 <\Gamma $ & $\mathcal{C}_E(VI_0)$ \\ \hline
 $\gamma=6/5$ & $\gamma<\Gamma$ & $\mathcal{S}(VI_0)$ \\ \hline
 $6/5<\gamma$ & $ \gamma<\Gamma$ & $\mathcal{E}(VI_0)$ \\ \hline

\end{tabular}
\caption{The future stable equilibrium points for various
equation of state parameters.}\label{tab:stable}
\end{table}


\section{An extreme tilted perfect fluid}
\label{sect:extreme} In this case we assume that the second fluid
is of extreme tilt; i.e., \beq \tilde{V}^2 = 1. \eeq This case was
not explicitly studied in \cite{hervik}.

By a careful analysis of the equilibrium points in the invariant
set $F(VI_0)$ we can prove the following:

\begin{thm} When $\tilde{V}^2 = 1$, all of the equilibrium
points of the resulting dynamical system of physical interest (in
the invariant set $F(VI_0)$) necessarily have the property that
(i) $V^2 = 0$ (i.e., the first fluid is comoving), (ii) $V^2 = 1$
(i.e., the first fluid is also of extreme tilt), or (iii)
$(v_1,v_2,v_3) = -\omega (\tilde{v}_1,\tilde{v}_2,\tilde{v}_3)$
(i.e., the two fluids are aligned).
\label{thm:41}\end{thm}

This makes the resulting analysis of the equilibrium points and
their stability relatively straightforward (and in many cases the
results can be inferred from the analysis of the single fluid case
\cite{hervik}).

For the extremely tilted invariant subspace, $\tilde{V}=1$, the
equation for $\tilde{\Omega}$ simplifies to (with $\Gamma<2$)
\beq
\tilde{\Omega}'=(2q-2-\tilde{\mathcal{S}})\tilde{\Omega}.
\label{eq:Extreme}\eeq 
The extremely tilted fluid loses its
dependence on the parameter $\Gamma$, and dynamically it behaves similarly to an anisotropic 
radiation fluid with $\Gamma=4/3$.\footnote{Compare eq.(\ref{eq:Extreme}) with a non-tilted radiation fluid with anisotropic stresses for which the evolution equation takes the form ${\Omega_r}'=(2q-2-\Sigma_{ab}\Pi^{ab}){\Omega}_r$  where $\Pi^{ab}$ is the expansion-normalised anisotropic stress tensor.}

Close to the initial singularity we typically have that $q\approx 2$
(separated by downward spikes in the chaotic case). Hence, since
$|\tilde{\mathcal{S}}|\leq 2\Sigma$, we see that $\tilde\Omega$ is
decaying towards zero as we approach the initial singularity. Thus
\emph{for $\Gamma<2$, the extremely tilted fluid will become
negligible close to the initial singularity (in the sense that $\tilde{\Omega}\rightarrow 0$).}

At late times we have that
$\tilde{\beta}\tilde{\Omega}\rightarrow0$. But since
$\tilde{\beta}=0$, the future evolution is more complicated; we
cannot conclude anything about the late-time behaviour from the
monotonic function alone. However, some general results can be
obtained from the existence of equilibrium points and their local
stability and using Theorem \ref{thm:41}.

From Table \ref{tab:stable} we see that when the second fluid is extreme, then the 
future attractors are: $\mathcal{I}_E(I)$, for $0<\gamma < 2/3$
and $4/3 <\Gamma $, $\mathcal{C}_E(VI_0)$ for $2/3<\gamma< 6/5$
and $4/3-(3\gamma-2)/12 <\Gamma $, and $\mathcal{E}(VI_0)$ for
$6/5<\gamma$ and $ \gamma<\Gamma$. But both $\mathcal{I}_E(I)$ and
$\mathcal{C}_E(VI_0)$ have $\tilde{\Omega}=0$ (and $\Omega \ne
0$), so that for $\gamma < 6/5$ the second (extreme) fluid is
dynamically negligible at late times. Only when $\gamma > 6/5$ can
the second fluid be non-negligible, and in this case the future
attractor is $\mathcal{E}(VI_0)$, where both fluids are of extreme
tilt.

In principle, we should study the special case in which both fluids
are of extreme tilt separately. However, the main results follow
directly from the arguments above. Both extreme
tilting fluids will be negligible dynamically at early times for
appropriate values of the parameters and the initial singularity
will be oscillatory in general. And at late times the local future
attractor will be $\mathcal{E}(VI_0)$ for $6/5 < \gamma < \Gamma$.

Thus we have the following: \emph{In the Bianchi type VI$_0$ model
the extremely tilted fluid becomes dynamically insignificant
($\tilde{\Omega}\rightarrow 0$) in the future if the non-extremely
tilted fluid has $\gamma<\frac 65$.} Note that this includes, for
example, the dust case (for which $\gamma=1$).

We have identified the extreme tilting fluid to be the second
fluid. In the above we assumed that $\gamma< \Gamma$. The case
$\gamma > \Gamma$ is not obtained by simply interchanging the two fluids in the case described above
(due to the asymmetry in the choice of the extreme fluid), and
care should be taken in describing the asymptotic dynamics in this
case. However, this can be done using Theorem \ref{thm:41} and the local
stability of the equilibrium points (see below).

\subsection{Null fluid; Relation to branes}

In brane-world cosmology matter fields and gauge interactions are
confined to a four-dimensional brane moving in a
higher-dimensional ``bulk" spacetime. The five-dimensional Weyl
tensor in the generalized Randall-Sundrum-type models
\cite{randall} is felt on the brane via its projection, ${\mathcal
E}_{\mu\nu}$. In general, in the 4-dimensional picture the
conservation equations do not determine all of the independent
components of ${\mathcal E}_{\mu\nu}$ on the brane. In these
models the gravitational field can also propagate in the extra
dimensions. For example, at sufficiently
high energies particle interactions can produce 5D gravitons which
are emitted into the bulk. In some applications these
gravitational waves may be approximated as of type N
\cite{coleyhervik}.

Let us assume that the 5-dimensional bulk is algebraically special
and of type N. Then there exists a null frame such that the 5-dimensional Weyl
tensor is given by
\cite{classification}: 
\beq
C_{abcd}=4C_{1i1j}\ell_{\{a}m^i_b\ell_cm^j_{d\}}. 
\eeq 
Defining
$n^c$ to be the normal vector on the brane, the non-local stresses
on the brane can be written as 
\beq
\mathcal{E}_{ab}=C_{1i1j}\left[\ell_a(m_c^in^c)-m^i_a(\ell_cn^c)\right]\left[\ell_b(m_c^jn^c)-m^j_b(\ell_cn^c)\right],
\eeq 
where $\mathcal{E}_{ab}n^b=\mathcal{E}^a_{~a}=0$. Using the
projection operator on the brane and defining $\hat{\ell}^b$ as the projection of the
null vector $\ell_b$ onto the brane, we consider the physically interesting case in which $\hat{\ell}^b\hat{\ell}_b=0$, so that $\hat{\ell}_a$ is \emph{null}.
The four-dimensional projected Weyl tensor on the brane can then be
written  \beq
\mathcal{E}_{\mu\nu}=-\left(\frac{\tilde{\kappa}}{\kappa}\right)^4\epsilon\hat{\ell}_{\mu}\hat{\ell}_{\nu}.
\eeq 
This is equivalent to the energy-momentum tensor of
a \emph{null} fluid, which is formally equivalent to the energy-momentum
tensor of an extreme tilted perfect fluid. Using a covariant
decomposition of $\mathcal{E}_{\mu\nu}$, the non-local energy
terms are given by: \beq
\mathcal{U}=\epsilon(\hat{\ell}_{\nu}u^{\nu})^2, ~~
\mathcal{Q}_{\mu}=\epsilon(\hat{\ell}_{\nu}u^{\nu})\hat{\ell}_{\mu},
~~\mathcal{P}_{\mu\nu}=\epsilon\hat{\ell}_{\langle\mu}\hat{\ell}_{\nu\rangle}.
\eeq The equations on the brane now close and the dynamical
behaviour can be analysed.

We can  investigate the effect this type N bulk may
have on the cosmological evolution of the brane. Since dynamically $\Gamma\sim 4/3$ in the isotropic case, for $\gamma>\Gamma=4/3$ we need to reverse the arguments given in
the previous subsection.  First, we note that if $\gamma > 4/3$,
then the null fluid is not dynamically important at early times;
that is, the effects of the projected Weyl tensor will not affect
the dynamical behaviour close to the initial singularity\footnote{In the brane scenario the Friedmann equation contains quadratic matter terms $\rho^2$ (compared to the general-relativistic $\rho$ dependence). so that close to the initial singularity we can effectively replace $\gamma\rightarrow 2\gamma$. This result is consequently consistent with  \cite{COLEY}.}. This
supports the result that an isotropic singularity is a local
stable past attractor, and hence constitute the most likely initial conditions in a classical brane-world \cite{COLEY}. Also, if $\gamma
< 4/3$, the extreme tilting fluid is dynamically negligible to
the future.

Therefore, the null brane fluid is not dynamically important
asymptotically at early and late times for all values of $\gamma$
of physical import, supporting the phenomenological analysis
presented in \cite{coleyhervik}.

\section{Stiff perfect fluid}
\label{sect:stiffone}

In this section we consider the case that the second fluid is a
stiff comoving fluid. Close to the initial singularity a scalar
field is essentially massless and can be modelled by a comoving
perfect fluid with a stiff equation of state parameter $\Gamma=2$.
The stiff fluid was not explicitly studied in \cite{hervik}. This will
enable us to consider the initial singular regime as
$\tau\rightarrow -\infty$ in this case. Unlike for
non-stiff fluids with $\Gamma<2$, for which the initial singular
regime possesses an oscillatory, and very likely a chaotic,
behaviour \cite{HBWII,WE,hbc}, in this case there exists a global past
attractor \cite{barrow:78}.

In the Appendix we have proven
\begin{par}\noindent\textbf{Theorem \ref{thm:51}}. \textit{ 
For $\gamma<\Gamma=2$, and $\tilde{\Omega}>0$ we have that}
\beq
\lim_{\tau\rightarrow-\infty}\tilde{V}=0, ~\lim_{\tau\rightarrow-\infty} \overline{\left(\Sigma^2+\tilde{\Omega}\right)}=1.
\label{eq:JacobsDisc}
\eeq\end{par}

The overbar denotes an appropriate mean value (see eq.(\ref{def:overbar})).
Hence, the stiff fluid will be asymptotically co-moving at early times. Note also that $\Sigma^2+\tilde{\Omega}=1$ implies $\Omega=N=0$. This means that the second fluid  $\gamma<\Gamma$ will be dynamically insignificant at early times.

For the stiff fluid case, $\Gamma=2$, there exist equilibrium
points given by (usually referred to as Jacobs disc):
\beq
\Sigma_+^2+\Sigma_{23}^2+\tilde{\Omega}=1, \quad \lambda=1, \nonumber \\
V=\tilde{V}=N=\Sigma_-=\Sigma_{12}=\Sigma_{13}=\Omega=0. \eeq
The
equilibrium points can thus be considered as the solid disc
 $\Sigma_+^2+\Sigma_{23}^2\leq 1$ in the
$(\Sigma_+,\Sigma_{23})$-plane. The centre of the disc represents
the FRW model with a stiff fluid and the unit circle correspond
to vacuum Kasner solutions.

\begin{figure}
  \includegraphics[width=5cm]{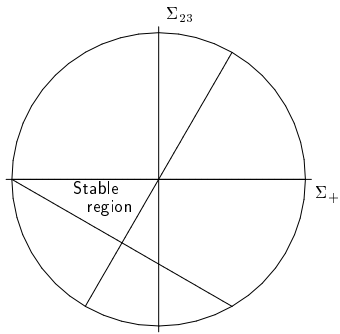}
  \caption{Jacobs Disc. Equilibrium points inside the triangular region are stable to the past.}\label{Fig:JD}
\end{figure}

By linearising the system of equations with respect to the Jacobs
disc, we get the following eigenvalues from the shear, $N$,
$\lambda$ and $\Omega$ equations: \beq \lambda_{1,2}=0, \quad
\lambda_{3}=-2\sqrt{3}\Sigma_{23}, \quad
\lambda_{4,5}=-3\Sigma_+\pm\sqrt{3}\Sigma_{23}\nonumber \\
\lambda_6=2(1+\Sigma_++\sqrt{3}\Sigma_{23}), \quad \lambda_7=-4\sqrt{3}\Sigma_{23},\quad
\lambda_8=3(2-\gamma).
 \eeq
The remaining (independent) eigenvalues come from the velocities $\tilde{v}_i$ which, according to Theorem \ref{thm:51}, have $\tilde{v}_i\rightarrow 0$.
Hence, there exists a triangular region inside the Jacobs disc
in which the equilibrium points are locally stable into the past (see Fig.\ref{Fig:JD}). This indicates that solutions including a stiff fluid are
not generically chaotic as we approach the initial singularity.
From the existence of the monotonic function $\chi$, we know that
the less stiff fluid has $\beta\Omega\rightarrow 0$ as we approach the 
initial singularity. From the
eq.(\ref{eq:JacobsDisc}) we see that $\tilde{\Omega}>0$ implies $\Omega\rightarrow 0$ at early times; i.e. we expect that the fluid with
$\gamma<2$ will decouple from the equations of motion and the
initial singularity is almost entirely determined by the stiffest
fluid. In this case the chaotic behaviour, conjectured to exist
for $\Gamma<2$, can be avoided.

Therefore to conclude: \emph{If one of the fluids is stiff
($\Gamma=2$), there exist local past attractors corresponding to
non-tilted solutions with a stiff perfect fluid. Furthermore, the less stiff fluid will be dynamically insignificant at early times.}

\section{Discussion}
\label{sect:disc}

We  have studied Bianchi type VI$_0$ universe models containing
two tilted $\gamma$-law perfect fluids using  dynamical systems
techniques. The state space is 11-dimensional. We utilized the
existence of a monotonic function, analysed the equilibrium points
in a particular important invariant subspace, derived Theorem 4.1,
and subsequently found  equilibrium points that are
future stable in the full 11-dimensional state space and are
consequently local attractors; the solutions, which are summarized
in Table 1, serve as late-time asymptotes for an open set of tilted
type VI$_0$ models containing two tilted perfect fluids.

We then studied the case in which one fluid is extremely tilting
in more detail (this analysis, plus the analysis of the stiff
fluid case, completed the analysis of \cite{hervik}). We
determined possible future behaviour and discussed when the
extreme fluid is negligible asymptotically.

We then applied these results to a problem in brane-world cosmology.
The four-dimensional equations on the brane are modified by the
effect of the projection of the five-dimensional Weyl tensor on
the brane, ${\mathcal E}_{\mu\nu}$, and when the bulk
gravitational field can be approximated as type N gravitational
waves close to the brane, this implies that the form of ${\mathcal
E}_{\mu\nu}$ is that of a null fluid, which is formally
dynamically equivalent to a perfect fluid with extreme tilt.
Therefore, we can model the dynamical effects of ${\mathcal
E}_{\mu\nu}$ in brane cosmology by assuming that the second fluid
is of extreme tilt. We found that the effects of ${\mathcal
E}_{\mu\nu}$ are not dynamically important (at least in this class
of models) in the asymptotic regime close to the singularity (as
expected from the work of \cite{coleyhervik}) nor to the future at
late times for physically important values for the fluid
parameters.

Finally, we studied the important case of a stiff fluid. We found
that in this case there is a local past attractor (5.2), so that
unlike previous analysis (for models with $\gamma<2$), there are
no chaotic oscillations as the initial singularity is approached.

\section*{Acknowledgments}
 This work was funded in part by The Research
Council of Norway, an Isaac Newton Studentship, and an AARMS
PostDoctoral Fellowship (SH).
\appendix
\section{Some simple proofs}

\begin{lem}
For $0< \gamma<\Gamma\leq 2$, we have
\beq
q\leq 2, \text{ and }~  q=2 \Rightarrow \Sigma^2+\tilde{\Omega}=1.
\label{eq:qineq}\eeq
\end{lem}
\begin{proof}
We can use the constraint equation and write
\[
q=2-2N^2-f(\gamma,V)\Omega-f(\Gamma,\tilde{V})\tilde{\Omega}\]
where
\beq
f(\gamma,V)\equiv\frac 12\frac{3(2-\gamma)+(5\gamma-6)V^2}{1+(\gamma-1)V^2}.
\eeq
Note that for $0<\gamma\leq 2$, $0\leq V\leq 1$, the function $f$ obeys $f(\gamma,V)\geq 0$, and equality holds if and only if $\gamma=2$, $V=0$. From this we can see that eq.(\ref{eq:qineq}) follows automatically.
\end{proof}

As the initial singularity is approached  the solutions typically 
become oscillatory and chaotic. In order to control this
oscillatory behaviour it is useful to introduce an appropriate
mean value, $\overline{A}$, defined by 
\beq
\overline{A}(\tau)\equiv \frac{1}{\tau_0-\tau}\int_{\tau}^{\tau_0}
A(t)dt, \quad \tau_0>\tau. 
\label{def:overbar}\eeq 
This definition of a mean value is
very useful since it basically measures the integrated effect of a
certain quantity. The possible sharp spikes appearing in the
curvature variables as the solution oscillates are therefore
integrated out. As $\tau\rightarrow -\infty$, these spikes can
therefore be ignored as an integrated effect. 

In the case of a stiff fluid, $\Gamma=2$, the following theorem is useful:

\begin{thm}\label{thm:51}
For $\gamma<\Gamma=2$, and $\tilde{\Omega}>0$ we have that
\[
\lim_{\tau\rightarrow-\infty}\tilde{V}=0, ~\lim_{\tau\rightarrow-\infty} \overline{\left(\Sigma^2+\tilde{\Omega}\right)}=1.
\]\end{thm}
\begin{proof}
For a stiff fluid the equations for $\tilde{V}$ and $\tilde{\Omega}$ are:
\beq
\tilde{V}'&=&(2-\tilde{\mathcal{S}})\tilde{V}, \nonumber \\
\tilde{\Omega}'&=& 2\left[(q-2)+(2-\tilde{\mathcal{S}})\frac{\tilde{V}^2}{1+\tilde{V}^2}\right]\tilde{\Omega}.
\label{eq:Gamma=2eq}\eeq
As shown above, $q$ is bounded by $q\leq 2$. Therefore,  there exists an $\epsilon(\tau)\geq 0$ such that   ${q}= 2-\epsilon$. Integrating the latter equation in (\ref{eq:Gamma=2eq}) (after dividing by $\tilde{\Omega}$ and replacing $(2-\tilde{\mathcal{S}})\tilde{V}$ with $\tilde{V}'$), we obtain ($\tau<\tau_0$)
\beq
\left.\ln\tilde{\Omega}\right|^{\tau_0}_{\tau}
=-2\int_{\tau}^{\tau_0}\epsilon d\tau+\left.\ln(1+\tilde{V}^2)\right|_{\tau}^{\tau_0}.
\eeq
Since $\ln(1+\tilde{V}^2)$ is bounded and $\epsilon\geq 0$, the right-hand side is bounded from above by $\ln 2$. The left-hand side, on the other hand, is bounded from below by $\ln\tilde{\Omega}(\tau_0)$ (which is negative). Hence, both sides must be bounded.  More precisely, there must exist a $\delta>0$ such that $\tilde{\Omega}\geq\delta >0$ for $\tau<\tau_0$. Moreover, boundedness implies that $\int_{\tau}^{\tau_0}\epsilon d\tau$ is bounded, and hence, 
\[ \bar{\epsilon}=\frac{1}{\tau_0-\tau}\int_{\tau}^{\tau_0}\epsilon d\tau \rightarrow 0\quad \text{for}\quad  \tau\rightarrow-\infty, \] 
which leads to $\overline{q}\rightarrow 2$ at early times. Using  eq.(\ref{eq:qineq}) we thus have $\lim_{\tau\rightarrow-\infty} \overline{\left(\Sigma^2+\tilde{\Omega}\right)}=1$.

Furthermore, we note that $\tilde{\mathcal{S}}$ obeys the bound $|\tilde{\mathcal{S}}|\leq 2\Sigma$, which for $0<\tilde{\Omega}\leq 1$ implies that $\tilde{V}$  is strictly monotonically increasing. Since $\tilde{V}$ is bounded, either $\tilde{V}\rightarrow 0$ or $\tilde{\mathcal{S}}\rightarrow 2$. The latter case can only happen if $\Sigma\rightarrow 1$, which implies that $\tilde{\Omega}\rightarrow 0$. However, this case is excluded since  $\tilde{\Omega}\geq \delta >0$ for $\tau<\tau_0$.  The theorem now follows.

\end{proof}

\section{Equilibrium points}
\label{sect:Eqpoints}
The system of equations possesses a wealth
of equilibrium points. Here, we will describe some of the
interesting ones which are important for the late-time
behaviour.
\subsection{Non-tilted + Non-tilted}
\subsubsection{$\mathcal{I}(I)$, FRW:}
$\Sigma^2=N=\lambda=0=V=\tilde{V}=\tilde{\Omega}=0$, $\Omega=1$,
$q=\frac 12(3\gamma-2)$, $0<\gamma,\Gamma <2$.
\paragraph{Eigenvalues:}
\beq \lambda_{1,2,3,4,5}=-\frac 32(2-\gamma), \quad
\lambda_{6,7}= \frac 12(3\gamma-2), \quad
\lambda_8=-3(\Gamma-\gamma), \quad
\lambda_{9,10,11}=(3\Gamma-4).\nonumber  \eeq

\subsubsection{$\mathcal{C}(VI_0)$, Collins VI$_0$:}
$\Sigma_-=\Sigma_{12}=\Sigma_{13}=\Sigma_{23}=\lambda=V=\tilde{V}=\tilde{\Omega}=0$,
$\Sigma_+=-\frac 14(3\gamma-2)$, $N^2=\frac
3{16}(3\gamma-2)(2-\gamma)$, $\Omega=\frac 34(2-\gamma)$,
$q=\frac 12(3\gamma-2)$, $2/3<\gamma <2$, $0<\Gamma <2$.
\paragraph{Eigenvalues:}
\beq \lambda_{1,2}=-\frac
34(2-\gamma)\left(1\pm\sqrt{5\gamma-6}\right), \quad
\lambda_{3,4}=-\frac
34(2-\gamma)\left(1\pm\sqrt{\frac{10-13\gamma}{2-\gamma}}\right),
\nonumber \\
\lambda_5=-\frac 32(2-\gamma), \quad \lambda_{6,7}=\frac
34(5\gamma-6), \quad \lambda_8=-3(\Gamma-\gamma), \nonumber \\
\lambda_9=(3\Gamma-4)+\frac 12(3\gamma-2), \quad
\lambda_{10,11}=(3\Gamma-4)-\frac 14(3\gamma-2)\nonumber  \eeq

\subsection{Non-tilted + Intermediately tilted}
\subsubsection{$\mathcal{I}_T(I)$, FRW with a tilted fluid:}
$\Sigma^2=N=\lambda=0=V=0$, $\Omega=1$, $q=\frac 12(3\gamma-2)$.
$0<\gamma<2$.

For the second fluid:
\[ \Gamma=\frac 43, ~\tilde{\Omega}=0,~\tilde{v}_1=0,~\tilde{v}_2=\tilde{v}_3=\frac{V}{\sqrt{2}}.\]
\paragraph{Eigenvalues:}
\beq \lambda_{1,2,3,4,5}=-\frac 32(2-\gamma), \quad
\lambda_{6,7}= \frac 12(3\gamma-2), \quad \lambda_8=(3\gamma-4),
\quad \lambda_{9,10,11}=0.\nonumber  \eeq

\subsubsection{$\mathcal{C}_T(VI_0)$, Collins type VI$_0$ with a tilted fluid:}
$\Sigma_-=\Sigma_{12}=\Sigma_{13}=\Sigma_{23}=\lambda=V=0$,
$\Sigma_+=-\frac 14(3\gamma-2)$, $N^2=\frac
3{16}(3\gamma-2)(2-\gamma)$, $\Omega=\frac 34(2-\gamma)$,
$q=\frac 12(3\gamma-2)$, $2/3<\gamma <2$.

For the second fluid:
\[ \Gamma=\frac 43-\frac 1{12}(3\gamma-2), ~\tilde{\Omega}=0,~\tilde{v}_1=0,~\tilde{v}_2=\tilde{v}_3=\frac{V}{\sqrt{2}}.\]

\paragraph{Eigenvalues:}
\beq \lambda_{1,2,3,4,5,6,7}~\text{ as for }~\mathcal{C}(VI_0),\quad\lambda_{8}=\frac{3\tilde{G}_-}{4\tilde{G}_+}(5\gamma-6),\quad \lambda_{9}=0, \nonumber \\
\lambda_{10,11}=-\frac32(3\gamma-2)\left(1\pm\sqrt{1-8\frac{(2-\gamma)V^2}{3\gamma-2}}\right)
\eeq
\subsection{Non-tilted + Extremely tilted}
 \subsubsection{$\mathcal{I}_E(I)$, FRW with an extremely tilted fluid:}
$\Sigma^2=N=\lambda=0=V=0$, $\Omega=1$, $q=\frac 12(3\gamma-2)$.
$0<\gamma<2$.

For the second fluid:
\[ 0<\Gamma<2, ~\tilde{\Omega}=0,~\tilde{v}_1=0,~\tilde{v}_2=\tilde{v}_3=\frac{1}{\sqrt{2}}.\]
\paragraph{Eigenvalues:}
\beq \lambda_{1,2,3,4,5}=-\frac 32(2-\gamma), \quad
\lambda_{6,7}= \frac 12(3\gamma-2), \nonumber \\
\lambda_8=(3\gamma-4), \quad
\lambda_{9}=-\frac{2(3\Gamma-4)}{2-\Gamma}, \quad
\lambda_{10,11}=0.\nonumber  \eeq The zero eigenvalues correspond
to the fact that the equilibrium point is a member of a
two-dimensional set of equilibrium points.

\subsubsection{$\mathcal{C}_E(VI_0)$, Collins type VI$_0$ with an extremely  tilted fluid:}
$\Sigma_-=\Sigma_{12}=\Sigma_{13}=\Sigma_{23}=\lambda=V=0$,
$\Sigma_+=-\frac 14(3\gamma-2)$, $N^2=\frac
3{16}(3\gamma-2)(2-\gamma)$, $\Omega=\frac 34(2-\gamma)$,
$q=\frac 12(3\gamma-2)$, $2/3<\gamma <2$.

For the second fluid:
\[ 0<\Gamma<2,~\tilde{\Omega}=0, ~\tilde{v}_1=0,~\tilde{v}_2=\tilde{v}_3=\frac{1}{\sqrt{2}}.\]

\paragraph{Eigenvalues:}
\beq \lambda_{1,2,3,4,5,6,7}~\text{ as for }~\mathcal{C}(VI_0),
\quad
\lambda_{8}=-\frac{3(2-\Gamma)}{4\Gamma}(5\gamma-6),\nonumber \\
\lambda_{9}=-\frac{2}{\Gamma}\left[3\Gamma-4+\frac
14(3\gamma-2)\right], \quad
\lambda_{10,11}=-\frac32(3\gamma-2)\left(1\pm\sqrt{\frac{11\gamma-18}{3\gamma-2}}\right)
\nonumber \eeq

\subsection{Intermediately tilted +  Extremely tilted}
\subsubsection{$\mathcal{S}(VI_0)$, a $\gamma=6/5$ bifurcation:}
\beq \gamma=\frac 65, \quad 0<\Gamma<2,\quad q=\frac 45, \quad
\Sigma_+=-\frac 25, \quad \Sigma_-=\Sigma_{23}=0, \quad
\Sigma_{13}=-\Sigma_{12}\nonumber \\
v_1=\tilde{v}_1=0, \quad v_2=v_3=\frac{V}{\sqrt{2}}, \quad 0\leq
V<1, \quad
\tilde{v}_2=\tilde{v}_3=\frac{1}{\sqrt{2}},\\
\Omega=\frac{\left[12-25(1+\lambda)N^2\right](5+V^2)}{50(1-V^2)},
\quad
\tilde{\Omega}=\frac{25N^2(2V^2+3\lambda+1)-6(5V^2+1)}{25(1-V^2)},
\nonumber \\
\Sigma_{12}=\frac{1}{10}\sqrt{25N^2(1-\lambda)-6}, \quad
N=\frac{\sqrt{B\pm 3(V+1)(1-\lambda)\sqrt{\sigma}}}{5\sqrt{A}}
 \eeq
 where
 \beq
 A&=& \frac{1+\lambda}{\lambda^2-4\lambda+7}\left\{\left[(\lambda^2-4\lambda+7)V+(1+\lambda)(\lambda-5)\right]^2-18(1-\lambda)^3\right\}
 \nonumber \\
 \sigma &=& \frac{\left[(\lambda^2-202\lambda+121)V+\lambda^2+98\lambda-11\right]^2-432\lambda(\lambda+5)(\lambda+1)}{\lambda^2-202\lambda+121}\nonumber \\
 B&=& \frac{3\left\{\left[(\lambda^2-2\lambda-39)V+\lambda^2+28\lambda+15\right]^2-36(3+8\lambda+23\lambda^2+2\lambda^3)\right\}}{\lambda^2-2\lambda-39} \nonumber \\
 \eeq
In addition we have to make sure that the energy densities are
positive and that $\Sigma_{12}$ is real. This puts bounds on $N$:
\beq \Omega\geq 0\quad
&\Rightarrow& \quad N^2\leq \frac{12}{25(1+\lambda)},\nonumber \\
\tilde{\Omega}\geq 0 \quad &\Rightarrow &\quad N^2\geq
\frac{6(5V^2+1)}{25(2V^2+3\lambda+1)},\nonumber \\
\Sigma_{12}^2\geq 0\quad &\Rightarrow &\quad N^2\geq
\frac{6}{25(1-\lambda)} \label{eq:BifurcBounds}\eeq The
bifurcation $\mathcal{S}(VI_0)$ is thus parameterized by
$\lambda$ and $V$ provided that the bounds
(\ref{eq:BifurcBounds}) hold.

The region is bounded by the following lines: \begin{enumerate}
\item{}$\tilde{\Omega}=0$: Given by the line bifurcation $\mathcal{L}(VI_0)$ for a single tilted fluid (see \cite{hervik}).
\item{} $\Omega=0$: \[ N^2=\frac{12}{25}, \quad \lambda=0.\]
\item{} $V=0$:
\[ N^2=\frac{9+42\lambda-3\lambda^2\pm 3(1-\lambda)\sqrt{\lambda^2-34\lambda+1}}{25(1+\lambda)(\lambda^2+14\lambda+1)},\quad
0\leq \lambda \leq 17-12\sqrt{2} \]
\item{} $V=1$:
\[ N^2=\frac{12}{25}, \quad \lambda=0, \quad \Omega+\tilde{\Omega}=\frac 6{25}. \]
\end{enumerate}
\paragraph{Eigenvalues:}
Using a combination of numerical and analytical work the
eigenvalues seem to be: \beq \lambda_{1,2}=0, \quad
\mathrm{Re}(\lambda_{3,4,5,6})=-\frac 35, \quad
\lambda_7=-\frac{6(5\Gamma-6)}{5(2-\Gamma)}, \nonumber \\
\lambda_8+\lambda_9=\lambda_{10}+\lambda_{11}=-\frac 65, \quad
\mathrm{Re}(\lambda_{8,9,10,11})<0. \nonumber  \eeq

\subsection{Extremely tilted + Extremely tilted}
\subsubsection{ $\mathcal{E}(VI_0)$: Extremely tilted type VI$_0$}
$V=\tilde{V}=1$, $v_1=\tilde{v}_1=0$,
$v_2=v_3=\tilde{v}_2=\tilde{v_3}=\frac{1}{2}{\sqrt{2}}$,
$\Sigma_-=\Sigma_{23}=\bar{N}=0$, $\Sigma_+=-\frac 25$,
$\Sigma_{12}=-\Sigma_{13}=\frac {\sqrt{6}}{10}$, $N=\frac
{2\sqrt{3}}5$, $\Omega+\tilde{\Omega}=\frac{6}{25}$,
$q=\frac{4}{5}$.

\paragraph{Eigenvalues:}
\beq  \lambda_1=-\frac{6(5\gamma-6)}{5(2-\gamma)},\quad
\lambda_{2,3}=-\frac 35\left(1\pm i\sqrt{19}\right),
\lambda_{4,5}=-\frac 35\left(1\pm i\sqrt{11}\right),\nonumber \\
\lambda_{6,7}=-\frac 35\left(1\pm i\sqrt{14+5\sqrt{2}}\right),
\quad \lambda_8=0, \quad
\lambda_9=-\frac{6(5\Gamma-6)}{5(2-\Gamma)},\nonumber \\
\lambda_{10,11}=-\frac 35\left(1\pm
i\sqrt{11+4\sqrt{2}}\right).\nonumber \eeq

\

\end{document}